\documentclass[twocolumn,twocolappendix]{aastex631}
\usepackage{graphicx}
\usepackage{txfonts}
\usepackage{subfigure}
\usepackage{threeparttable}
\usepackage{tabularx}
\usepackage{natbib}
\usepackage{multirow}
\usepackage{makecell}
\usepackage{ulem}
\usepackage{booktabs} 

\usepackage{amsmath}

\begin{document}

\title{Analysis of the \textit{Gaia} Data Release 3 parallax bias at bright magnitudes}

\author{Ye Ding}
\affiliation{Shanghai Astronomical Observatory, Chinese Academy of Sciences, Shanghai 200030, China}
\affiliation{University of Chinese Academy of Sciences, Beijing 100049, China}

\author{Shilong Liao}
\affiliation{Shanghai Astronomical Observatory, Chinese Academy of Sciences, Shanghai 200030, China}
\affiliation{University of Chinese Academy of Sciences, Beijing 100049, China}

\author{Shangyu Wen}
\affiliation{Shanghai Astronomical Observatory, Chinese Academy of Sciences, Shanghai 200030, China}
\affiliation{University of Chinese Academy of Sciences, Beijing 100049, China}

\author{Zhaoxiang Qi}
\affiliation{Shanghai Astronomical Observatory, Chinese Academy of Sciences, Shanghai 200030, China}
\affiliation{University of Chinese Academy of Sciences, Beijing 100049, China}

\correspondingauthor{Shilong Liao; Zhaoxiang Qi}
\email{shilongliao@shao.ac.cn; zxqi@shao.ac.cn}



\begin{abstract}
The combination of visual and spectroscopic orbits in binary systems enables precise distance measurements without additional assumptions, making them ideal for examining the parallax zero-point offset (PZPO) at bright magnitudes ($G < 13$) in \textit{Gaia}. We compiled 249 orbital parallaxes from 246 binary systems and used Markov Chain Monte Carlo (MCMC) simulations to exclude binaries where orbital motion significantly impacts parallaxes. After removing systems with substantial parallax errors, large discrepancies between orbital and \textit{Gaia} parallaxes, and selecting systems with orbital periods under 100 days, a final sample of 44 binaries was retained.The weighted mean PZPO for this sample is -38.9 $\pm$ 10.3 $\mu$as, compared to -58.0 $\pm$ 10.1 $\mu$as for the remaining systems, suggesting that orbital motion significantly affects parallax measurements. These formal uncertainties of the PZPO appear to be underestimated by a factor of approximately 2.0.  For bright stars with independent trigonometric parallaxes from VLBI and \textit{HST}, the weighted mean PZPOs are -14.8 $\pm$ 10.6 and -31.9 $\pm$ 14.1 $\mu$as, respectively. Stars with $G \leq 8$ exhibit a more pronounced parallax bias, with some targets showing unusually large deviations, likely due to systematic calibration errors in \textit{Gaia} for bright stars. The orbital parallaxes dataset compiled in this work serves as a vital resource for validating parallaxes in future \textit{Gaia} data releases.

\end{abstract}



 \section{Introduction} \label{sec:intro}

      
    \textit{Gaia} Data Release 3 (GDR3) contains results for more than 1.8 billion sources in the magnitude range $G=$ 3-21 \citep{2021A&A...649A...1G}. High-precision parallaxes are crucial for advancing our understanding of the structure of the Milky Way. The precise parallax measurements of bright stars are particularly valuable for determining the three-dimensional positions and velocities of nearby stars, enabling detailed studies of the solar neighborhood \citep{2021A&A...649A...6G}. However, systematic errors exist in the parallax data in \textit{Gaia}. A more concerning issue in the GDR3 astrometric solutions \citep{2021A&A...649A...2L} is the presence of complex variations in the mean along-scan (AL) residual as a function of magnitude for stars with $G < 13$ (where $G$ is the magnitude in the \textit{Gaia} passband). These variations arise from the use of gates to prevent pixel saturation. For stars with $G \lesssim 8$, the residual increases significantly due to partial image saturation, adding complexity to the parallax bias for bright stars.
    
    Systematic errors exist in the astrometric data in \textit{Gaia}, which have been studied by many researchers. \citeauthor{2021A&A...649A...4L} (\citeyear{2021A&A...649A...4L}, hereafter L21) presented a global parallax zero-point offset (PZPO) of -17 $\mu as$, based on quasars (QSOs) in GDR3. 
    Additionally, several studies have investigated the PZPO of GDR3, covering partial range in $G<13$ , based on different tracers. The tracers include eclipsing binary stars which span $5 \lesssim G \lesssim 12$ ($\sim$ -37 $\pm$ 20 $\mu$as, \citealt{2021ApJ...907L..33S}), red clump stars ranging from $G$ $\simeq$ 10 to 16 ($\sim$ -26 $\mu as$, \citealt{2021ApJ...910L...5H}), red giant branch stars in the range $G$ $\simeq$ 9 to 13 ($\sim$ -22 $\mu as$, \citealt{2021AJ....161..214Z}), giant stars spanning $G$ $\simeq$ 9 to 16 ($\sim$ -27.9 $\mu as$, \citealt{2022AJ....163..149W}), Cepheids spanning $6 \lesssim G \lesssim 11$ ($\sim$ -14 $\pm$ 6 $\mu as$, \citealt{2021ApJ...908L...6R}), and stars in open clusters ranging from $G$ $\simeq$ 8 to 17 ($\sim$ -59 $\pm$ 20 $\mu as$, \citealt{2021A&A...649A...5F}).
    Furthermore, additional studies have focused on the PZPO at fainter magnitudes (\citealt{2021PASP..133i4501L}; \citealt{2021ApJ...911L..20R}; \citealt{2021ApJ...909..200B}).
    With the exception of the QSOs that truly have a parallax of zero, other objects with derived parallaxes are based on direct or indirect assumptions, such as a period-luminosity relation, an absolute magnitude, asteroseismology, or the presumption that all stars within a cluster share the same parallax. 
    
    However, knowledge of both the visual and the spectroscopic orbits in binary systems enables precise distance to be determined, without any additional assumptions \citep{2020MNRAS.492.2709P}.
    \cite{2023A&A...669A...4G} (hereafter G23) compiled a list of 192 orbital parallax determinations for 186 systems, based on known orbital elements of binaries, to investigate the PZPO of GDR3 at bright magnitudes. They presented the weighted average of -41.7 $\mu as$, based on their standard selection (SS) of binaries, which removed objects with large parallax errors, unrealistically large differences between the orbital and \textit{Gaia} parallaxes, and large goodness-of-fit (GOF) parameters. After restricting these parameters more strictly, the value for the final selection (FS) was refined to -26.7 $\mu as$.


    The aim of this work is to have an independent investigation into the PZPO of GDR3 at bright magnitudes. To this end, we compile a comprehensive dataset of binaries with derived orbital parallaxes from various studies. Our sample also encompasses the complete set from G23. In our analysis, we will simulate the binaries with known orbital elements, to remove the orbital effect in the PZPO. What's more, we also use the bright stars with independent parallaxes determined from other surveys, including Very Long Baseline Interferometry (VLBI) astrometry \citep{2019ApJ...875..114X} and \textit{Hubble} Space Telescope (HST; \citealt{2021A&A...654A..20G}),  to enhance our investigation of the PZPO at bright magnitudes.
    
        
    The paper is structured as follows. Section \ref{sec:sample} details the samples of binaries with orbital parallaxes and stars with measured parallaxes from VLBI and HST. In Section \ref{sec:results}, we present the PZPOs of GDR3 at bright magnitudes, based on the samples described in Section \ref{sec:sample}. Section \ref{sec:discuss} discusses the PZPOs derived from different samples and visualizes various studies of the PZPO. Section \ref{sec:summary} concludes the paper. 

    \begin{table*}
    \caption{Sample of stars (selected entries)}           
    \label{sample}      
    \centering  
    \begin{threeparttable}
    \setlength{\tabcolsep}{2pt} 
    \begin{tabularx}{\textwidth}{c c c c c c c c c c c c} 
    \hline\hline  
    WDS & HD & HIP & P & T & a & e & i  & $\omega$ & $\Omega$ & $\pi_{o}$  & Reference  \\
        &    &     & (d) & (mjd) & (mas)  &  & (deg) & (deg) & (deg) &(mas)  & \\
    \hline  
    
    00369+3343 & 3369 & 2912 & 143.53 & 58705.0279 & 6.69 & 0.542 & 103.0 & 170.7 & 94.7 & 3.61 &
    {\cite{2020MNRAS.492.2709P}} \\
    &  &  &  $\pm$ 0.06 & $\pm$ 0.4 & $\pm$ 0.05 & $\pm$ 0.006  &  $\pm$ 0.2 & $\pm$ 0.7 & $\pm$ 0.2 & $\pm$ 0.03  &\\
    
    00490+1656 & 4676 & 3810 & 13.8245 & 50905.9746 & 6.55 & 0.2366 & 73.92 & 203.057 & 207.41 & 43.50 & {\cite{2005ApJ...626..431K}} \\
     &  &  &  $\pm$ 0.000043 & $\pm$ 0.0067 & $\pm$ 0.01 & $\pm$ 0.0006 & $\pm$ 0.80 & $\pm$ 0.073 & $\pm$ 0.65 & $\pm$ 0.09   &\\
    
    00572+2325 & 5516 & 4463 & 115.72 & 58753.8869 & 10.37 & 0.006 & 30.5 & 215.0 & 69.4 & 13.37 & {\cite{2020MNRAS.492.2709P}} \\
     &  &  & $\pm$ 0.01 & $\pm$ 1.0958 & $\pm$ 0.03 & $\pm$ 0.002 & $\pm$ 0.4 & $\pm$ 4.0 & $\pm$ 0.6 & $\pm$ 0.17 & \\
    
    01028+3148 & 6118 & 4889 & 81.1262 & 31308.153 & 5.56 & 0.8956  & 143.40 & 346.6 & 167.8 & 8.86  & {\cite{2004ApJ...610..443K}} \\
     &  &  & $\pm$ 0.0003 & $\pm$ 0.023 & $\pm$ 0.04 & $\pm$ 0.0020 & $\pm$ 1.3 & $\pm$ 2.0 & $\pm$ 1.7 & $\pm$ 0.07 & \\
    
    01108+6747 & 6840 & 5531 & 2722.0 & 52622.2 & 83.0 & 0.7442 & 52.0 & 215.2 & 151.8 & 16.64  & {\cite{2023A&A...669A...4G}} \\
     &  &  & $\pm$ 1.0 & $\pm$ 0.9 & $\pm$ 12.45 & $\pm$ 0.0020 & $\pm$ 12.0 & $\pm$ 0.5 & -- & $\pm$ 2.63 &\\
    
    01237+3743 & 8374 & 6514 & 35.36836 & 54292.7022 & 5.05 & 0.6476 & 140.64 & 145.18 & 336.2 & 16.21  & {\cite{2020AJ....160...58L}} \\
     &  &  & $\pm$ 0.00005 & $\pm$ 0.004 & $\pm$ 0.02 & $\pm$ 0.0005 & $\pm$ 0.45 & $\pm$ 0.1 & $\pm$ 1.2 & $\pm$ 0.07  &\\
    
    01277+4524 & 8799 & 6813 & 254.900 & 54214.835 & 38.0 & 0.142  & 62.49 & 278.87 & 115.94 & 39.12  & {\cite{2014AJ....148...48F}} \\
     &  &  & $\pm$ 0.196 & $\pm$ 3.187 & $\pm$ 1.0 & $\pm$ 0.012 & $\pm$ 2.1 & $\pm$ 2.01 & $\pm$ 4.38 & $\pm$ 1.97  &\\
    
    01321+1657 & 9312 & 7143 & 36.5192 & 56614.154 & 4.85 & 0.1433 & 103.4 & 203.386 & 237.0 & 17.56  & {\cite{2023A&A...672A.119G}} \\
     &  &  & $\pm$ 0.00002 & $\pm$ 0.004 & $\pm$ 0.03 & $\pm$ 0.0001 & $\pm$ 0.2 & $\pm$ 0.041 & $\pm$ 0.3 & $\pm$ 0.1  &\\
    
    01374+2510 & 9939 & 7564 & 25.20896 & 51786.408 & 4.944 & 0.102 & 61.56 & 313.07 & 262.3 & 23.68  & {\cite{2006ApJ...644.1193B}} \\
     &  &  & $\pm$ 0.00007 & $\pm$ 0.063 & $\pm$ 0.018 & $\pm$ 0.001 & $\pm$ 0.25 & $\pm$ 0.88 & $\pm$ 0.2 & $\pm$ 0.12  & \\
    
    01376-0924 & 10009 & 7580 & 10518.97536 & 47862.39203 & 324.0 & 0.7980  & 96.60 & 251.6 & 159.60 & 27.0  & {\cite{2000A&AS..145..215P}} \\
     &  &  & $\pm$ 281.23649 & $\pm$ 4.38291 & $\pm$ 5.4 & $\pm$ 0.0066 & $\pm$ 0.33 & $\pm$ 0.67 & $\pm$ 0.73 & $\pm$ 1.0 & \\
    
    01379-8259 & 10800 & 7601 & 638.6662 & 58818.902 & 78.23 & 0.1912 & 47.60 & 151.16 & 296.0 & 37.11  & {\cite{2020MNRAS.492.2709P}} \\
     &  &  & $\pm$ 0.0005 & $\pm$ 0.0026 & $\pm$ 0.47 & $\pm$ 0.0018 & $\pm$ 0.48 & $\pm$ 0.58 & $\pm$ 0.3 & $\pm$ 0.37  & \\
    
    01437+5041 & 10516 & 8068 & 126.6982 & 56109.53 & 5.89 & 0.0  & 77.6 & 0 & 295.7 & 5.38  & {\cite{2015A&A...577A..51M}} \\
     &  &  & $\pm$ 0.0035 & $\pm$ 0.08 & $\pm$ 0.02 & $\pm$ 0.1 & $\pm$ 0.3 & -- & $\pm$ 0.3 & $\pm$ 0.09  & \\
    
    02057-2423 & 12889 & 9774 & 943.055 & 55531.705 & 48.0 & 0.769 & 75.9 & 287.9 & 127.2 & 22.33  & {\cite{2014AJ....147..123T}} \\
     &  &  & $\pm$ 1.096 & $\pm$ 5.479 & $\pm$ 3.0 & $\pm$ 0.015 & $\pm$ 4.0 & $\pm$ 1.5 & $\pm$ 4.6 & $\pm$ 1.46  & \\
    
    02124+3018 & 13480 & 10280 & 14.7302 & 53352.5 & 2.1 & 0.0035 & 58.0 & 29.0 & -- & 11.83  & {\cite{2023A&A...669A...4G}} \\
     &  &  & $\pm$ 0.0009 & $\pm$ 2.7 & $\pm$ 0.9 & $\pm$ 0.0042 & $\pm$ 4.0 & -- & -- & $\pm$ 5.07  &\\
    
    02128-0224 & 13612 & 10305 & 94.786 & 45645.659 & 13.98 & 0.6920  & 25.6 & 76.23 & 240.3 & 26.29  & {\cite{2022AJ....163..118A}} \\
     &  &  & $\pm$ 0.004 & $\pm$ 0.099 & $\pm$ 0.75 & $\pm$ 0.0027 & $\pm$ 7.8 & $\pm$ 0.41 & $\pm$ 1.4 & $\pm$ 0.21  &\\
    
    02171+3413 & 13974 & 10644 & 10.0195 & 58817.966 & 9.80 & 0.0107 & 163.0 & 171.9 & 37.0 & 139.09  & {\cite{2020MNRAS.492.2709P}} \\
     &  &  & $\pm$ 0.0002 & $\pm$ 0.7793 & $\pm$ 0.13 & $\pm$ 0.0055 & $\pm$ 3.6 & $\pm$ 29.0 & $\pm$ 1.9 & $\pm$ 28.77  & \\
    
    02211+4246 &  & 10952 & 116147.020 & 47599.418 & 0.94 & 0.802 & 147.0 & 175.0 & 158.0 & 17.0  & {\cite{2000A&AS..145..215P}} \\
     &  &  & $\pm$ 6209.117 & $\pm$ 51.134 & $\pm$ 0.02 & $\pm$ 0.007 & $\pm$ 2.3 & $\pm$ 3.7 & $\pm$ 3.5 & $\pm$ 1.4 & \\
    
    02262+3428 & 15013 & 11352 & 2533.0 & 57300.5 & 99.0 & 0.300 & 49.9 & 0.0 & 16.0 & 22.04  & {\cite{2020MNRAS.492.2709P}} \\
     &  &  & $\pm$ 5.0 & $\pm$ 9.0  & -- & $\pm$ 0.005 & -- & $\pm$ 2.0 & -- & $\pm$ 0.40 &  \\
    
    02422+4012 & 16739 & 12623 & 330.982 & 49111.918 & 53.18 & 0.6574 & 128.17 & 49.29 & 269.29 & 41.19  & {\cite{2006AJ....131.2695B}} \\
     &  &  & -- & -- & $\pm$ 0.15 & -- & $\pm$ 0.14 & -- & -- & $\pm$ 0.21  &\\
    
    02442-2530 & 17134 & 12780 & 2443.2619 & 56508.134 & 100.2 & 0.4999  & 42.00 & 191.2 & 183.77 & 22.25  & {\cite{2020MNRAS.492.2709P}}\\
     &  &  & $\pm$ 1.8628 & $\pm$ 0.007 & $\pm$ 0.6 & $\pm$ 0.0015 & $\pm$ 0.72 & $\pm$ 0.59 & $\pm$ 0.49 & $\pm$ 0.35 & \\
     
    02539-4436 & 18198 & 13498 & 18795.364 & 54188.6 & 296.7 & 0.774  & 49.5 & 16.2 & 298.3 & 14.8  & {\cite{2016AJ....152..138T}}\\
     &  &  & $\pm$ 299.499 & $\pm$ 51.1 & $\pm$ 4.4 & $\pm$ 0.013 & $\pm$ 1.9 & $\pm$ 1.7 & $\pm$ 1.3 & $\pm$ 0.8  & \\
    \hline               
    \end{tabularx}
    \begin{tablenotes}
    \footnotesize
        \item Notes. Column 1: WDS number. Column 2: HD number. Column 3: HIP number. Column 4: Period with error in days. Column 5: Time passage through periastron with error in mjd. Column 6: Semi-major axis with error in mas. Column 7: Eccentricity with error. Column 8: Inclination with error in degrees. Column 9: The argument of pericenter with error in degrees. Column 10: The longitude of the ascending node with error in degrees. Column 11: Orbital parallax with error in mas. Column 12: References. 
        The table presented here includes only a subset of the columns. The complete table, with additional columns, is detailed in Appendix \ref{appendix:a}.
    \end{tablenotes} 
    \end{threeparttable}
    \end{table*}

\section{Sample} \label{sec:sample}

    \subsection{Binaries}\label{sec:sample-binaries}

            We compile a list of 249 orbital parallaxes for 246 systems from various literature, with GDR3 identifiers provided by SIMBAD. Table \ref{sample} lists the adopted orbital elements and orbital parallaxes, along with the relevant references, for a total of 246 systems. For three binaries, several components have been resolved (WDS 03272+0944, 20396+0458, and 22388+4419), and therefore the table has 249 entries. Among them, only one system ROXs 47A have no WDS, HD, and HIP number. Compared to G23, our sample contains 67 new binaries and 157 overlapped binaries but with orbital parallaxes obtained from other studies.
            Figure \ref{fig:compare_tw_g23} shows the distribution of the magnitude for binaries from this work (TW), G23, and overlapped sample, respectively. Among 249 sources, only 35 have a non-zero NSS flag in GDR3. In other words, 86\% of the known binaries have not been identified as binary in GDR3, likely due to GDR3 focusing on the most significant systems. 
            
            To investigate the PZPO, quality filtering is required. For 249 entries, we apply the following criteria:
            
            
           \begin{figure}
                \centering
                \subfigure{\includegraphics[width=0.8\linewidth]{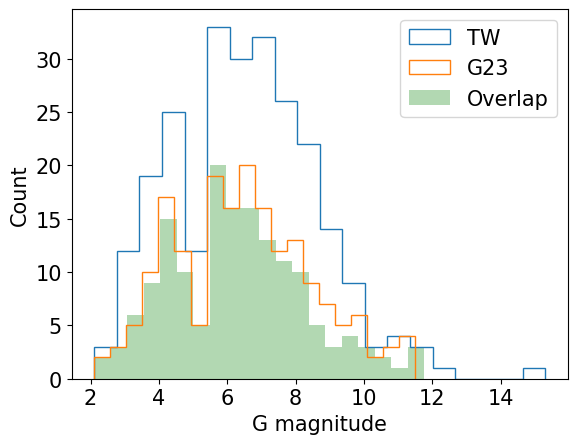}}
                \caption{ Histogram of the magnitude for binaries from this work (TW, 249), G23 (186), and overlapped sample (157). } 
                \label{fig:compare_tw_g23}
            \end{figure}

            \begin{align}\label{eq:filtering}
                \left\{     
                    \begin{aligned}
                        (i) \quad &astrometric\_params\_solved = 31 \quad or \quad 95, \\
                        (ii) \quad &\sigma_{\pi_{Orb}} < 5 * \sigma_{\pi_{GDR3}},  \\
                        (iii) \quad &\sigma_{\pi_{Orb}}, \sigma_{\pi_{GDR3}} < 2  mas \\
                        (iv) \quad & \left |  (\pi_{GDR3} - \pi_{Orb}) \right | / \sqrt{{{\sigma }_{\pi_{GDR3}}}^{2}+{{\sigma }_{\pi_{Orb}}}^{2}} <5
                    \end{aligned}
                \right.
            \end{align}
            
            Criterion (i) selects the sources that have five- or six-parameter astrometric solutions in GDR3; criterion (ii) selects the sources where the error in the orbital parallax should be smaller than five times the error in the GDR3 parallax; criterion (iii) selects the sources where the error in the orbital and GDR3 parallax should be smaller than 2 mas; and criterion (iv) selects the sources where the absolute parallax difference should be zero within five times the formal uncertainty. Among 249 sources, 132 survived the filtering by Eq. (\ref{eq:filtering}), including 103 and 29 sources with five- and six- parameter astrometric solutions, respectively. 
    
        \begin{figure}
            \centering
            \subfigure{\includegraphics[width=0.8\linewidth]{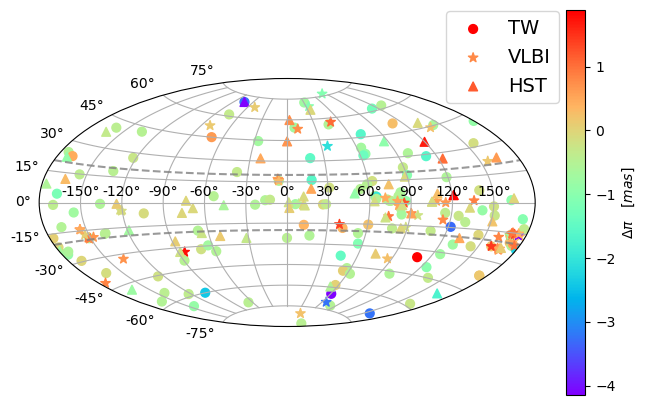}}
            \caption{Map of the difference ($\Delta \pi$) of GDR3 parallax minus external parallax in Galactic coordinate for the filtered samples from this work (TW, 132), VLBI (64), and HST (55), respectively. The dotted lines represent the Galactic latitude $\pm 20^{\circ}$.} 
            \label{fig:compare_pzpo}
        \end{figure}
    
    \subsection{Stars from VLBI and HST}\label{sec:sample-VLBI,HST}
        Bright stars with independent parallaxes determined from other surveys are considered as well. First, the VLBI astrometry is capable of measuring parallaxes with accuracies of $\sim$ 10 $\mu as$ \citep{2014ARA&A..52..339R}, which has an astrometric quality comparable to that of GDR3. We cross-match 108 VLBI sources of \cite{2019ApJ...875..114X} with GDR3 at a radius of 100 mas, retaining 102 unique matches. Among 102 matches, 2 have a non-zero NSS flag. What's more, 44 are identified as binary from \cite{2019ApJ...875..114X}. Only 1 is identified as binary both from \textit{Gaia} and \cite{2019ApJ...875..114X} (AR Lac). \cite{2021A&A...654A..20G} compiled a list of 111 objects with the independent trigonometric parallax data from HST, who provide GDR3 identifiers in their table. Of the 111 objects, only 1 is not listed in GDR3 (Polaris A), and 8 have no parallax listed in GDR3. Among the left 102 objects, 5 have a non-zero NSS flag. 
        
        For VLBI and HST sources, we apply the following criteria:
        \begin{align}\label{eq:filtering_vlbi}
            \left\{     
                \begin{aligned}
                    (i) \quad &astrometric\_params\_solved = 31 \quad or \quad 95, \\
                    (ii) \quad &G < 13,  \\
                    (iii) \quad & \left | \Delta \pi \right | / \sigma_{\Delta} < 5
                \end{aligned}
            \right.
        \end{align}
        Criterion (i) selects the sources that have five- or six-parameter astrometric solutions in GDR3; criterion (ii) selects the sources brighter than $G$ = 13; and criterion (iii) selects the sources where the absolute difference between the GDR3 and external parallax should be zero within five times the formal uncertainty. 64 VLBI sources survived the filtering by Eq. \ref{eq:filtering_vlbi}, including 51 and 13 sources with five- and six- parameter astrometric solutions, respectively. 
        55 HST sources survived the filtering, including 39 and 16 sources with five- and six- parameter astrometric solutions, respectively. 
        
        Figure \ref{fig:compare_pzpo} shows the map of the difference ($\Delta \pi$) of GDR3 parallax minus external parallax in Galactic coordinate for the filtered samples from TW, VLBI, and HST, respectively. In total, 59, 43, and 31 sources from TW, VLBI, and HST are located within the Galactic plane ($\left | b\right | \leqslant {20}^{\circ}$), respectively.

        \begin{figure}
            \centering
            \subfigure{\includegraphics[width=0.8\linewidth]{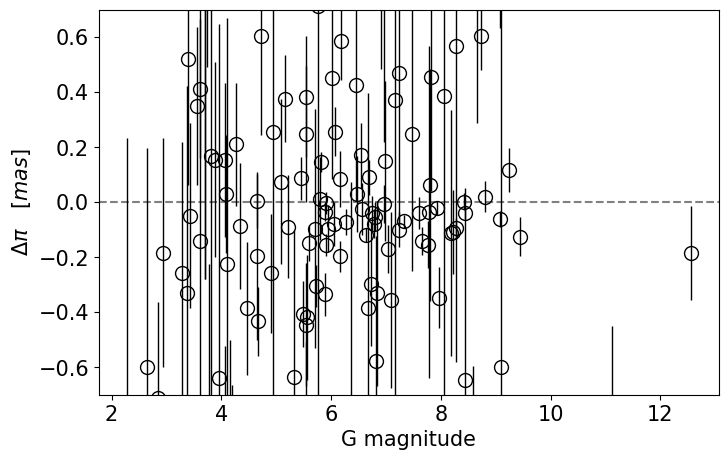}}
            \subfigure{\includegraphics[width=0.8\linewidth]{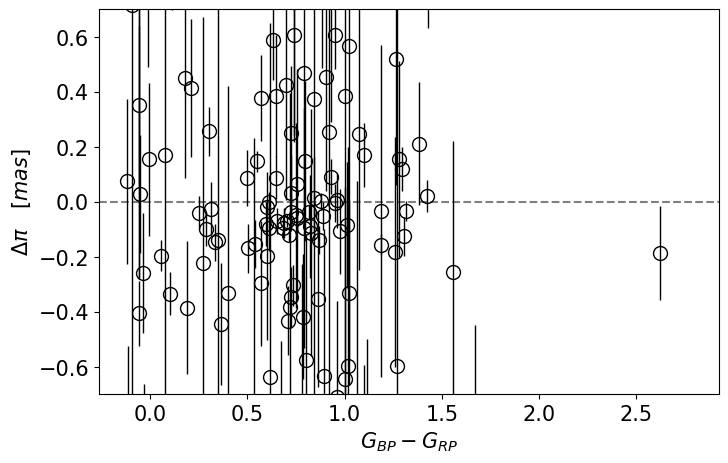}}
            \subfigure{\includegraphics[width=0.8\linewidth]{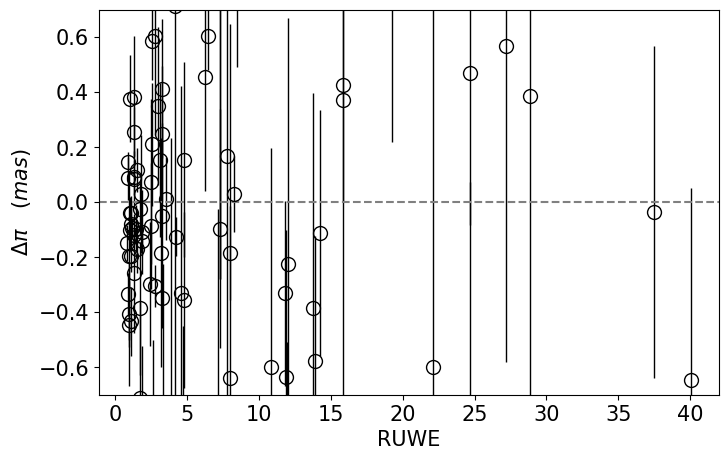}}
            \caption{Parallax difference of GDR3 parallax minus orbital parallax for 132 filtered binaries, plotted against magnitude, Bp-Rp color, and RUWE. Black circles show the parallax differences of individual filtered sources.} 
            \label{fig:pzpo_binary}
        \end{figure}

\section{Results} \label{sec:results}


    \subsection{Binaries} \label{subsec:binaries}
        
        Figure \ref{fig:pzpo_binary} shows the parallax difference of GDR3 parallax minus orbital parallax for the filtered binaries (132), plotted against magnitude, Bp-Rp color, and renormalised unit weight error (RUWE), respectively. The magnitude ranges from $G$ $\simeq$ 2.28 to 12.55. 65 are brighter than $G$ = 6. The Bp-Rp color spans from $G_{BP}-G_{RP}$ $\simeq$ -0.12 to 2.78, with a median value of 0.75. The RUWE value, indicating the goodness-of-fit of the single-star model to the observations in the astrometric solution (ideally=1.0; \citealt{2021A&A...649A...2L}), ranges from 0.82 to 40.0. 14 and 35 sources have a RUWE less than 1.0 and 1.4. The global weighted mean PZPO is -48.8 $\pm$ 7.2 $\mu as$. However, the PZPO of GDR3 contains an additional effect arising from the orbital motion of the binary star. Thus, we have to consider the orbital effect.

        To this end, we simulate binaries with known orbital elements to generate new parallaxes using the Markov chain Monte Carlo (MCMC). We applied 10 000 times MCMC samples on each binary by sampling the parameters within 3 times errors. For more details see Appenix \ref{sec:simulation}. 
        We define a solution as `good' if each binary's solutions satisfy the condition $\left | \pi_{GDR3}-\pi_{Simu} \right | / \pi_{GDR3} < 0.2$ with a confidence level of 95\% in MCMC, where $\pi_{GDR3}$ and $\pi_{Simu}$ indicate the GDR3 and simulated parallax, respectively. Among 132 filtered binaries, 93 are identified as `good' solutions. 
        
\begin{figure}
        \centering
        \subfigure{\includegraphics[width=0.8\linewidth]{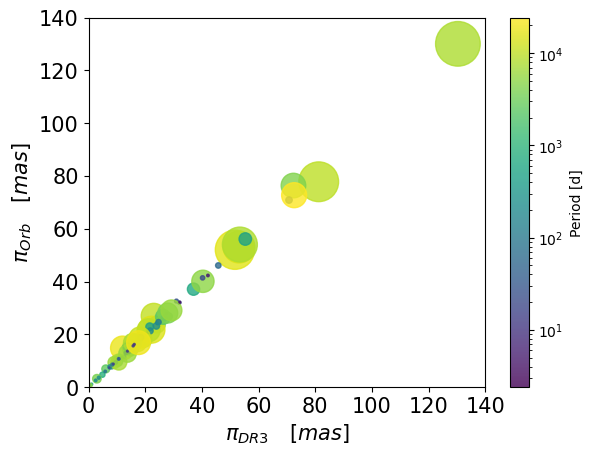}}
        \caption{Orbital parallax for the `good' binaries (93) plotted against GDR3 parallax. The size of each data point is proportional to the semi-major axis, while the color of the points corresponds to the orbital period. } 
        \label{fig:oplx_vs_gplx_goodsol}
        \end{figure}

        We try to minimize the orbital effect in the PZPO by selecting binaries whose parallaxes are insignificantly affected by orbital motions. Thus, we further focus on analyzing 93 `good' binaries. Figure \ref{fig:oplx_vs_gplx_goodsol} shows the orbital parallax for the 93 `good' binaries plotted against the GDR3 parallax. The period range from 2.39 to 24 392 days, and the semi-major axis range from 0.79 to 1032.00 $mas$. Binaries with wider orbits tend to have longer periods, and vice versa. 
        The NSS solutions with short orbital periods ($P$ $\lesssim$ 100 days) are hard to solve because of the insignificant orbital motion of binaries \citep{2023A&A...674A..10H}. To exclude objects significantly influenced by orbital motions, we select a final subset of 44 binaries with orbital periods shorter than 100 days.
        
\begin{figure}
        \centering
        \subfigure{\includegraphics[width=0.8\linewidth]{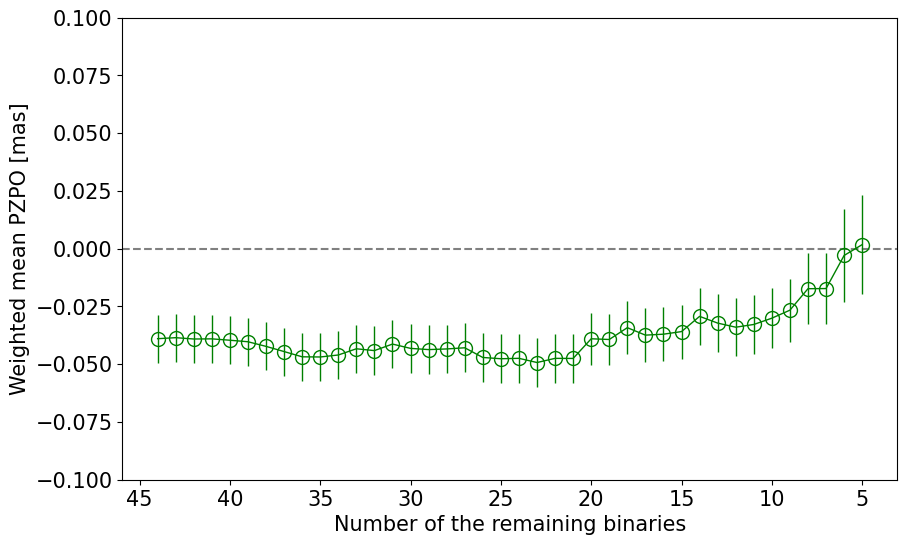}}
        \subfigure{\includegraphics[width=0.8\linewidth]{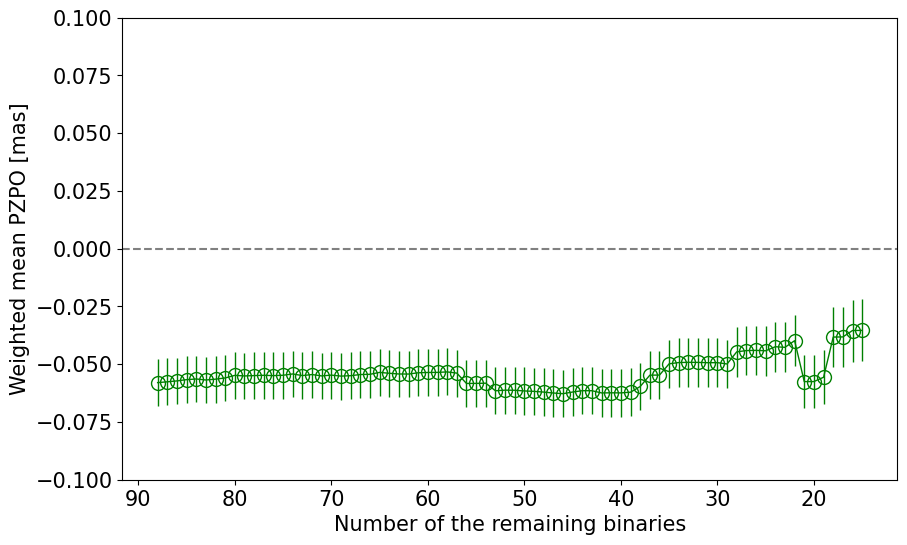}}
        \caption{Weighted mean PZPO plotted against the number of the final selection and remaining binaries. At each step, the source with the largest absolute value of the parallax difference is removed. The gray dashed line represents zero value.} 
        \label{fig:pzpo_size_binary}
        \end{figure}

        Figure \ref{fig:pzpo_size_binary} shows the weighted mean PZPO plotted against the number of the final selection and remaining binaries. We rank the 44 and remaining 88 binaries based on their absolute values of the parallax differences, iteratively removing the source with the largest value. At each step, the weighted mean PZPOs are recalculated for the remaining sample, continuing until the sample sizes are reduced to 5 and 15, respectively. The PZPO for the final selection and remaining sources are relatively stable on the whole. We assume that the orbital effect in the parallax can be negligible for the final selection. The weighted mean PZPO for this final selection is -38.9 $\pm$ 10.3 $\mu as$. Additionally, for the remaining 88 binaries in the filtered sample, the weighted mean PZPO is -58.0 $\pm$ 10.1 $\mu as$. The results of parallax bias from these two subsets are significantly different, highlighting the importance of accounting for and filtering out the effects of orbital motion on parallax measurements.

    \subsection{Stars from VLBI and HST}\label{subsec:vlbi}

        We also investigate the PZPO for the filtered stars from VLBI and HST described in Section \ref{sec:sample-VLBI,HST}.
        First, the global weighted mean PZPO for VLBI sources (64) is  -15.6 $\pm$ 9.3 $\mu as$. Among the 64 sources, 32 are single stars while the remaining 32 are identified as binary from \cite{2019ApJ...875..114X} or \textit{Gaia}. We supplement the orbital periods for 23 of the 32 binaries. Similar to Section \ref{subsec:binaries}, 13 binaries with orbital periods shorter than 100 days are selected. We combine 13 binary stars with 32 single stars as the final selected sample. The weighted mean PZPO for the final selection is -14.8 $\pm$ 10.6 $\mu as$.
        
        In addition to the VLBI sources, the global weighted mean PZPO for HST sources (55) is  -34.5 $\pm$ 14.0 $\mu as$. Among 55 filtered HST sources, 3 with a non-zero NSS flag are removed. The weighted mean PZPO for the final selection (52) is -31.9 $\pm$ 14.1 $\mu as$.

        \begin{table*}
        \centering
            \caption{Weighted mean PZPO derived from the binaries, VLBI, and HST sources, respectively.}
            \label{table:pzpo}      
            \setlength{\tabcolsep}{2pt} 
            \begin{threeparttable}
                 \begin{tabular}{c| c c c| c c c| c c c }     
                    \hline\hline          
                    & \multicolumn{3}{c|}{Binary}  & \multicolumn{3}{c|}{VLBI} & \multicolumn{3}{c}{HST} \\
                    \cline{2-10}
                    & $\Delta \pi$ & $G$ & N & $\Delta \pi$ & $G$ & N & $\Delta \pi$ & $G$ & N \\ 
                    & ($\mu as$) & ($mag$) & & ($\mu as$) & ($mag$) & &  ($\mu as$) & ($mag$) &  \\
                    \hline
                    Filtered sources & -48.8 $\pm$ 7.2 & 6.11 & 132  & -15.6 $\pm$9.3 & 9.07 & 64 & -34.5 $\pm$ 14.0 & 8.34 & 55 \\ 
                    Filtered sources for $G \leq 8$ & -56.7$\pm$8.6 & 5.54 & 111 & -60.6$\pm$28.0 & 6.33 & 23 & -24.6$\pm$27.0 & 5.24 & 21 \\
                    Filtered sources for $G > 8$ & -28.8$\pm$13.5 & 9.12 & 21 & -10.0$\pm$9.9 & 10.61 & 41 & -38.1$\pm$16.4 & 10.26 & 34 \\
                    \hline
                    
                    Final selection & -38.9 $\pm$ 10.3 & 6.00 & 44 & -14.8$\pm$10.6 & 8.58 & 45 & -31.9$\pm$14.1  & 8.20 & 52 \\
                    Final selection for $G \leq 8$ & -49.9$\pm$15.5 & 5.27 & 35 & -53.4$\pm$28.1 & 6.33 & 20 & -24.6$\pm$27.0 & 5.24 & 21 \\
                    Final selection for $G > 8$ & -30.1$\pm$13.9 & 8.84 & 9 & -8.5$\pm$11.4 & 10.39 & 25 & -34.6$\pm$16.5 & 10.20 & 31 \\
                    Supplemented final selection & +7.9$\pm$10.0 & 5.98 & 48 & -- & -- & -- & --  & -- & -- \\
                    \hline
                    
                    Remaining sources & -58.0 $\pm$ 10.1 & 6.16 & 88 & -18.6$\pm$20.1 & 10.24 & 19 & -410.2$\pm$169.9 & 10.91 & 3 \\
                    Remaining sources for $G \leq 8$ & -59.7$\pm$10.3 & 5.66 & 76 & -766.8$\pm$277.9 & 6.34 & 3 & -- & -- & -- \\
                    Remaining sources for $G > 8$ & -7.9$\pm$55.7 & 9.33 & 12 & -14.6$\pm$20.1 & 10.97 & 16 & -- & -- & -- \\
                    Supplemented remaining sources & -83.4$\pm$10.0 & 6.20 & 97 & -- & -- & -- & -- & -- & -- \\
                  \hline                                   
                \end{tabular}
                \begin{tablenotes}[flushleft]
                \footnotesize
                    \item Notes. Results are given for the binary (Cols. 2-4), VLBI (Cols. 5-7), and HST (Cols. 8-10), respectively, and give the weighted mean parallax difference with its formal uncertainty, the average G-mag of the objects, and  the number of the objects considered.
                \end{tablenotes} 
            \end{threeparttable}
        \end{table*}

            \begin{figure}
            \centering
            \subfigure{\includegraphics[width=0.8\linewidth]{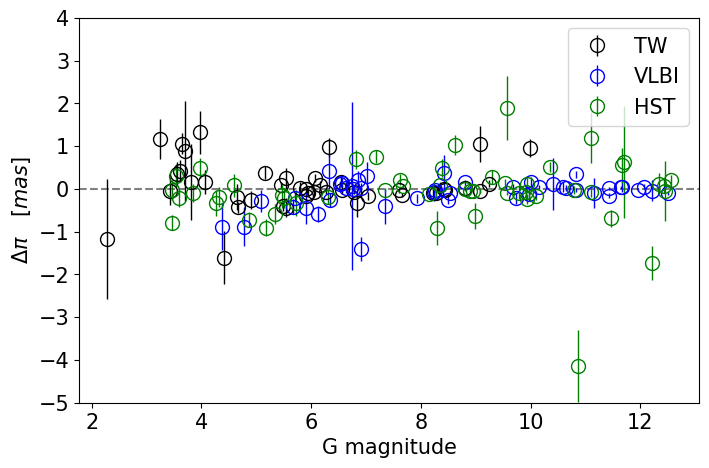}}
            \subfigure{\includegraphics[width=0.8\linewidth]{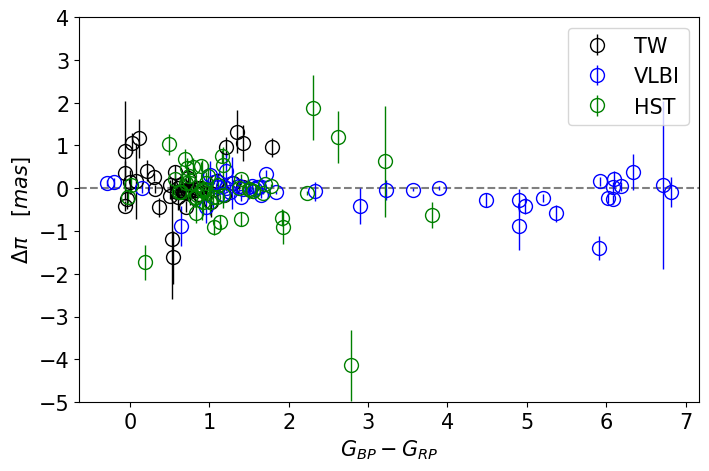}}
            \subfigure{\includegraphics[width=0.8\linewidth]{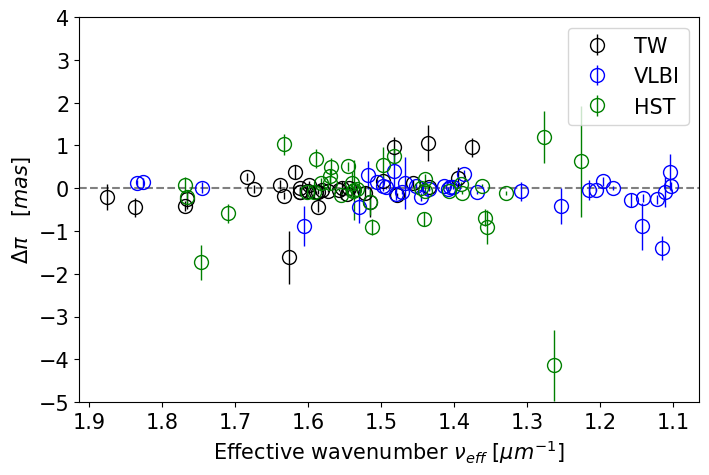}}
            \caption{Difference of GDR3 parallax minus external parallax ($\Delta \pi$) plotted against magnitude, Bp-Rp, and effective wavenumber for the final selection of this work (TW, 44), VLBI (45), and HST (52), respectively.} 
            \label{fig:compare_color}
        \end{figure}  
        
\section{Discussion} \label{sec:discuss}

    \subsection{PZPO}\label{sec:discuss-pzpo}
        In Section \ref{sec:results}, we present the PZPOs at bright magnitudes ($G<$ 13) in GDR3, derived from different samples. 
        Figure \ref{fig:compare_color} shows the difference of GDR3 parallax minus external parallax ($\Delta \pi$) plotted against magnitude, Bp-Rp color, and effective wavenumber for the final selection of TW, VLBI, and HST, respectively. 35, 20, and 21 sources from binaries, VLBI, and HST are brighter than $G$ = 8. The Bp-Rp color distribution of the binaries is concentrated, with a maximum value of 1.79, while other samples exhibit a broader and more dispersed distribution. The median effective wavenumber is 1.59, 1.41, and 1.52 $\mu m^{-1}$ for the final selection for TW, VLBI, and HST, respectively. 
        
        Table \ref{table:pzpo} lists the weighted mean PZPO derived from the binaries, VLBI, and HST sources, respectively. In Section \ref{sec:sample-binaries}, we excluded the binaries whose absolute parallax differences greater than five times the formal uncertainty, removing 13 sources (HD 27149, WDS 00369+3343, 00490+1656, 01374+2510, 01418+4237, 04306+1542, 04506+1505, 05025+4105, 05407-0157, 11557+1539, 21041+0300, 22070+2521, and 22375+3923).  
        Now we also list the results of binaries supplementing these sources in Table \ref{table:pzpo}. Among these sources, HD 27149, WDS 00490+1656, 01374+2510, and 22070+2521 belong to the final selection subset, while the left 9 binaries are in the remaining sources subset. The parallax difference for HD 27149 is 1.09 $\pm$ 0.04 $\mu$as with a relatively large weight, primarily leading to a strange positive PZPO (+7.9 $\mu as$) for the supplemented final selection.
        In the results for the final selection, the parallax differences are closer to zero-point than values for the filtered sources. 
        Comparing results for the final selection in different magnitude ranges, the parallax differences for binaries and HST with $G \le 8$ are closer to the zero-point, while the value for VLBI in this range is significantly larger. The binaries are generally brighter than the VLBI and HST sources, showing the contribution of this work to the study of the PZPO at brighter magnitudes ($G \leq 8$). Targets with magnitude $G \le 8$ exhibit a more significant parallax bias. Furthermore, some individual targets with high parallax accuracy show unusually large and significant biases (for example, HD 27149). This could result from systematic errors caused by calibration issues in \textit{Gaia}’s instruments for bright-end targets.
        
\begin{figure}
            \centering
            \subfigure{\includegraphics[width=0.8\linewidth]{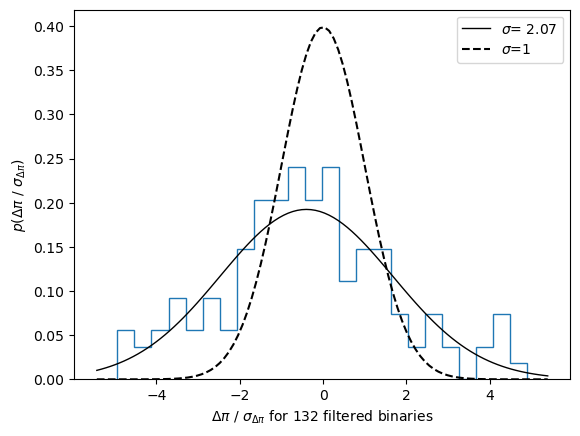}}
            \caption{Distributions of uncertainty normalized parallax difference $\Delta \pi / \sigma_{\Delta} = (\pi_{GDR3} - \pi_{Orb})| / \sqrt{{{\sigma }_{\pi_{GDR3}}}^{2}+{{\sigma }_{\pi_{Orb}}}^{2}}$ for 132 filtered binaries. The quantity would be expected to follow a Gaussian distribution with $\sigma$ = 1 (dotted lines) if the formal parallax uncertainties were accurate. $\sigma>$  1 points towards underestimated parallax uncertainties. Solid black line shows the Gaussian fit, with $\sigma$ = 2.07.} 
            \label{fig:gauss_fit}
        \end{figure}

        However, the astrometric uncertainties are underestimated in GDR3, especially for bright binaries (\citealt{2021MNRAS.506.2269E}, hereafter EB21). To address this, we attempt to estimate the uncertainties of the PZPO using various methods. Following Figure 15 in EB21, we plotted the distribution of $\Delta\pi / \sigma_{\Delta\pi}$ for the 132 filtered binaries, along with the corresponding fitted Gaussian function, as shown in Figure \ref{fig:gauss_fit}. The best-fitting $\sigma$, representing the uncertainty underestimation factor, was determined to be 2.07. Through bootstrap re-sampling, the median underestimation factor is estimated to be approximately 1.80. Additionally, Figure 16 in EB21 indicated an underestimation factor ranging from 2.1 to 2.8 for binaries with $7 < G < 12$ and $ruwe > 1.4$. These three results are consistent, suggesting that the underestimated factor of the PZPO's uncertainty for our binary sample is approximately 2.0. Consequently, the corrected uncertainties of the PZPO for the 132 filtered, 44 final, and 88 remaining binaries are estimated to be around 14.4, 20.6, and 20.2 $\mu as$, respectively. The upcoming \textit{Gaia} data release could potentially resolve this issue and provide a more precise estimation of the uncertainty.

    \subsection{Other studies on the PZPO}\label{sec:discuss-other}
    
        Systematic errors are inevitable in the published astrometric data in \textit{Gaia} \citep{2021A&A...649A...2L}. High-precision and high-accuracy parallaxes are essential for various fields in astronomy, leading to the investigation and correction of the PZPO essential. L21 proposed a recipe to correct the PZPO of GDR3, which is a function of G-band magnitude, colour information and ecliptic latitude of the sources. 
        However, \cite{2024A&A...691A..81D} found that the L21 correction did not apply to the Galactic plane effectively, and presented a correction applicable specifically to the Galactic plane, which offered improvements over the L21 correction.

\begin{figure}
            \centering
            \subfigure{\includegraphics[width=0.8\linewidth]{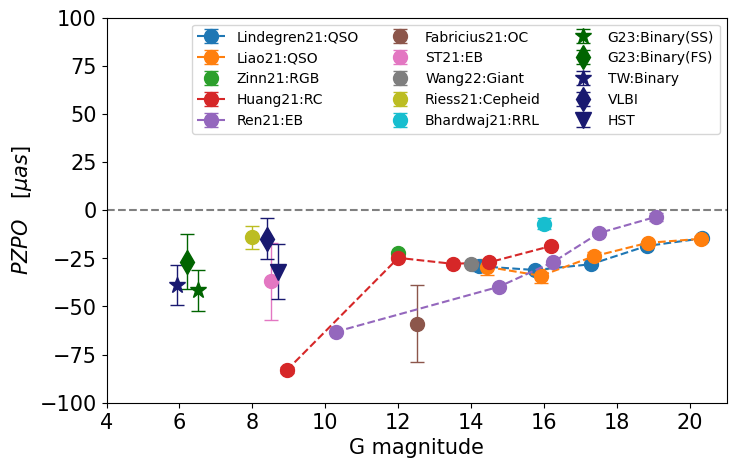}}
            \caption{The PZPO from various studies and the final selection in this work, plotted against the median magnitude of the objects used in each study.} 
            \label{fig:pzpo_studies}
        \end{figure}

        In addition to the corrections to the PZPO, various studies investigating the PZPO are summarized here. Section \ref{sec:intro} introduces studies focusing on the bright magnitude range, while additional research focus on the faint end in GDR3. \cite{2021PASP..133i4501L} found a PZPO of -21 $\mu as$ using 299 004 confirmed external QSOs , together with quasars identified in GDR3. The magnitudes of their objects range from $G$ $\simeq$ 14 to 21. \cite{2021ApJ...911L..20R} reported a global PZPO of -28.6 $\pm$ 0.6 $\mu$as based on 110 000 W Ursae Majoris (EW)-type eclipsing binary systems, with magnitudes ranging from $G$ $\simeq$ 13 to 19. \cite{2021ApJ...909..200B} reported a median PZPO of -7 $\pm$ 3 $\mu as$ based on 400 Galactic RR Lyrae stars within the brightness range $15 \lesssim G \lesssim 17$.
        Figure \ref{fig:pzpo_studies} visualizes the PZPOs of GDR3 presented by various studies and this work. \cite{2021A&A...649A...5F} derived the average PZPOs of -59 $\pm$ 20 $\mu as$ and -91 $\pm$ 20 $\mu as$ from open clusters in \cite{2013A&A...558A..53K} and \cite{2014A&A...564A..79D}, and we visualize the result of the open cluster in \cite{2013A&A...558A..53K}, with clusters spanning magnitudes of $8 \lesssim G \lesssim 17$.
        With the exception of G23 and our work, other studies used the various objects with derived parallaxes based on direct or indirect assumptions. 
        Compared to G23, our result for the final selected binaries extends the coverage to brighter magnitudes.
        
\section{Summary} \label{sec:summary}

    Knowledge of both the visual and the spectroscopic orbits in binary systems enables precise distance to be determined, offering a new method of independently investigating the PZPO at bright magnitudes. We investigate the improved parallaxes of the NSS solutions from the astrometric binary pipeline in GDR3 and found that the PZPO contains an orbital effect.
    In this work, we compile a list of 249 orbital parallaxes for 246 systems from various literature.  Binaries whose parallaxes are virtually unaffected by orbital motion are retained. We simulate binaries with known orbital elements using the MCMC to remove the orbital effect in the PZPO, while G23 removed the orbital effect primarily by filtering of good-of-fitness statistics in GDR3 (ipd\_gof\_harmonic\_amplitude). Except for derived orbital parallaxes, independent parallaxes determined from VLBI and HST are considered as well. The weighted mean PZPOs and their formal uncertainties are -38.9 $\pm$ 10.3, -14.8 $\pm$ 10.6, and -31.9 $\pm$ 14.1 $\mu as$ for the final selection of binaries, VLBI, and HST, respectively. The formal uncertainties of the PZPO for our binary sample are underestimated by approximately 2.0. Our work contributes the study of the PZPO at the brightest magnitude. Further progress is expected with \textit{Gaia} DR4, which will have improved parallax uncertainties and reduced systematic errors. Targets with a magnitude of $G \leq 8$ exhibit a notably larger parallax bias. Some individual targets, despite their high parallax accuracy, exhibit unusually large and notable biases. These deviations may result from systematic errors linked to calibration issues in \textit{Gaia}'s instruments for bright-end targets. Consequently, caution is recommended when utilizing parallax data for \textit{Gaia} targets, especially for those brighter than 8th magnitude. Orbital parallaxes are among the few methods that can directly assess parallax bias for bright sources (G $\leq$ 13). This study presents the most comprehensive compilation of orbital parallax measurements to date, gathered from a wide range of literature. The value of this dataset extends beyond GDR3, serving as a critical resource for validating parallaxes in future \textit{Gaia} data releases.


\section*{Acknowledgments}


This work has made use of data from the European Space Agency (ESA) mission \textit{Gaia} (\href{https://www. cosmos.esa.int/gaia}{https://www.cosmos.esa.int/gaia}), processed by the \textit{Gaia} Data Processing and Analysis Consortium (DPAC, \href{https://www.cosmos.esa.int/web/gaia/dpac/consortium}{https://www.cosmos.esa. int/web/gaia/dpac/consortium}). Funding for the DPAC has been provided by national institutions, in particular the institutions participating in the \textit{Gaia} Multilateral Agreement. We are also very grateful to the developers of the TOPCAT (\citealt{2005ASPC..347...29T}) software.
 This work has been supported by the National Natural Science Foundation of China (NSFC) through grants 12173069, the Strategic Priority Research Program of the Chinese Academy of Sciences, Grant No.XDA0350205, the Youth Innovation Promotion Association CAS with Certificate Number 2022259, the Talent Plan of Shanghai Branch, Chinese Academy of Sciences with No.CASSHB-QNPD-2023-016. We acknowledge the science research grants from the China Manned Space Project with NO. CMS-CSST-2021-A12 and NO.CMS-CSST-2021-B10.

%






\bibliography{ref}{}
\bibliographystyle{aasjournal}

\appendix

\setcounter{table}{0}   
\renewcommand{\thetable}{A.\arabic{table}} 

\section{The details of our binary sample}\label{appendix:a}

In Section \ref{sec:sample-binaries}, we compile a list of 249 orbital parallaxes for 246 systems drawn from various literature sources. Table \ref{sample} provides the orbital elements and parallaxes for 21 binaries, representing a subset of the full dataset and only covering selected columns. For the complete binary sample, additional columns include the semi-amplitudes of the binary components, a flag indicating whether the binary is included in G23, the orbital parallax from G23 for binaries included in G23, a flag for `good' solutions as described in Section \ref{subsec:binaries}, and the matched GDR3 source id with associated information. Detailed descriptions of all columns in the binary sample are provided in Table \ref{tab:appendix}.

\begin{table*}
    \centering  
    \caption{Details columns of the binary sample}           
    \label{tab:appendix} 
    \setlength{\tabcolsep}{0.5pt}
    \begin{tabular}{l l l l l} 
    
    \hline\hline  
    Column & Name & Unit & Description & Example value \\
    \hline  
    
    1 & WDS & -- & WDS number & 00490+1656 \\
    2 & HD & -- & HD number & 4676 \\
    3 & HIP & -- & Hipparcos id & 3810 \\
    4 & dr3\_id & -- & GDR3 source id & 2781872793183604096 \\
    5 & Per & day & Orbital period & 13.8245 \\
    6 & e\_Per & day & Orbital period error & 0.000043 \\
    7 & T & mjd & The time passage through periastron & 50905.9746 \\
    8 & e\_T & mjd & The time passage through periastron error & 0.0067 \\
    9 & amaj & mas & Semi-major axis & 6.55 \\
    10 & e\_amaj & mas & Semi-major axis error & 0.01 \\
    11 & ecc & -- & Eccentricity & 0.2366 \\
    12 & e\_ecc & -- & Eccentricity error & 0.0006 \\
    13 & incl & degree & Inclination & 73.92 \\
    14 & e\_incl & degree & Inclination error & 0.80 \\
    15 & omega & degree & The argument of periastron & 203.057 \\
    16 & e\_omega & degree & The argument of periastron error & 0.073 \\
    17 & OMEGA & degree & The position angle of the ascending node & 207.41 \\
    18 & e\_OMEGA & degree & The position angle of the ascending node error & 0.65 \\
    19 & $K_{1}$ & km/s & Semi-amplitude of component 1 & 57.552 \\
    20 & $e\_K_{1}$ & km/s & Semi-amplitude of component 1 error & 0.037 \\
    21 & $K_{2}$ & km/s & Semi-amplitude of component 2 & 59.557 \\
    22 & $e\_K_{2}$ & km/s & Semi-amplitude of component 2 error & 0.038 \\
    23 & oplx & mas & Orbital parallax & 43.50 \\
    24 & e\_oplx & mas & Orbital parallax error & 0.09 \\
    25 & oplx\_G23 & mas & Orbital parallax from G23 & 43.29 \\
    26 & e\_oplx\_G23 & mas & Orbital parallax error from G23 & 0.44 \\
    27 & parallax & mas & GDR3 parallax & 42.75 \\
    28 & parallax\_error & mas & GDR3 parallax error & 0.11 \\
    29 & flag & -- & Indicate whether the binary is included in G23: & B1 \\
	  & & & `A' means the new binary not included in G23; & \\
	  & & & `B1' means that G23 gives a different orbital parallax than the one used here adopted from reference; & \\
	  & & & `B2' mean that the orbital parallax in G23 was adopted. & \\
    30 & good\_sol & -- & True if the filtered binary's solutions satisfy the condition $\left | \pi_{GDR3}-\pi_{Simu} \right | / \pi_{GDR3} < 0.2$ \\
        & & & with a confidence level of 95\% in MCMC & False \\
    31 & Ref & -- & Reference & Konacki05 \\
    32 & BibCode & -- & BibCode & 2005ApJ...626..431K \\
    \hline            
    \end{tabular}
    \end{table*}

\section{Simulation procedure}\label{sec:simulation}



The simulation time period for GDR3 observations spans from J2014.5624599 TCB to J2017.4041495 TCB, equivalent to approximately 2014-07-25T10:30:00 UTC to 2017-05-28T08:45:00 UTC \citep{2021A&A...649A...1G}. We utilize the \textit{Gaia} Observation Forecast Tool\footnote{\href{https://gaia.esac.esa.int/gost/}{https://gaia.esac.esa.int/gost/}} to predict the transit times $t_{i}$ for each source crossing \textit{Gaia}'s focal plane, along with the associated observational scan angles $\theta_{i}$ and parallax factors $\Pi_{i}$.


The displacement of a source on the celestial tangent plane projected to the along\textendash scan (AL) direction, $\eta$, is described by \citep{2014ApJ...797...14P,2022A&A...665A.111W}:
\begin{align}
    \begin{split}
        \eta(t_{i})=&\left[\Delta\alpha^{\star}+\mu_{\alpha^{\star}}\left(t_{i}-t_{0}\right)+BX+GY\right]\sin\theta_{i}\\
        &+\left[\Delta\delta+\mu_{\delta}\left(t_{i}-t_{0}\right)+AX+FY\right]\cos\theta_{i}\\
        &+\Pi_{i}\varpi,
    \end{split}
    \label{eq:double_star_model}
\end{align}
where $\Delta\alpha^{\star}\equiv\Delta\alpha\cos\delta,\mu_{\alpha^{\star}}\equiv\mu_{\alpha}\cos\delta$, and $A,B,F,G$ are Thiele\textendash Innes elements defined as follows \citep{wright2009efficient}:
\begin{align}
    A&\equiv a(\cos\omega\cos\Omega-\sin\omega\sin\Omega\cos i),\\
    B&\equiv a(\cos\omega\sin\Omega+\sin\omega\cos\Omega\cos i),\\
    F&\equiv a(-\sin\omega\cos\Omega-\cos\omega\sin\Omega\cos i),\\
    G&\equiv a(-\sin\omega\sin\Omega+\cos\omega\cos\Omega\cos i),
\end{align}
and the elliptical rectangular coordinates are
\begin{align}
    X&=\cos E-e,\\
    Y&=\sqrt{1-e^{2}}\sin E,
\end{align}
where the eccentric anomaly $E$ is obtained from solving the Keplerian equation
\begin{align}
    \begin{split}
        E-e\sin E&=\frac{2\pi}{P}\left(t-t_{0}\right)\\
        &=\frac{2\pi}{P}t+M_{0}.
    \end{split}
\end{align}
Additionally, the AL astrometric error $\sigma$ is calculated by Eq. (21) in \cite{2021MNRAS.502.1908E}. Assuming the error is Gaussian, the final mock observation result is 
\begin{align}
    \eta_{\text{sim}}(t_{i})=\eta(t_{i})+\text{GaussianRandom}(0,\sigma^{2}),\label{eq:simulation}
\end{align}
where the subscript `sim' denotes that Eq. \eqref{eq:simulation} is only used in the simulation.

We fitted the mock data using single star model. The single star model is the binary model Eq. \eqref{eq:double_star_model} without the Keplerian part
\begin{align}\label{eq:single_star_model}
    \eta_{\text{fit}}(t_{i})=\left[\Delta\alpha^{\star}+\mu_{\alpha^{\star}}\left(t_{i}-t_{0}\right)\right]\sin\theta_{i}
    +\left[\Delta\delta+\mu_{\delta}\left(t_{i}-t_{0}\right)\right]\cos\theta_{i}+\Pi_{i}\varpi.
\end{align}
Then the log\textendash likelihood is
\begin{align}
    \ln L=-\frac{1}{2}\sum_{i=1}^{N_{\text{obs}}}\left[\frac{\eta_{\text{fit}}\left(t_{i}\right)-\eta_{\text{obs}}\left(t_{i}\right)}{\sigma}\right]^{2}.
\end{align}

We applied Markov Chain Monte Carlo (MCMC) algorithm provided by \texttt{emcee} package \citep{2013PASP..125..306F} to sample the log-likelihood function 10 000 times per source.  For every sampler, we generated a new set of initial values based on the astrometric parameters with errors from the dataset, then slightly turbulated the initial values to produce 64 walkers for each parameter. Each sampler walked until it converged, or up to 20 000 steps, including 1 000 steps as burn\textendash in.





\end{document}